\documentclass{mem}
\usepackage{natbib}\usepackage{txfonts}\usepackage{balance}
\usepackage{graphicx}
\usepackage[a4paper]{hyperref}
\idline{79}{001}
\def\kms{km ${\rm s}^{-1}$}

\def\ch2{$\chi^2$}

\def\kms {\hbox{${\rm km\ s}^{-1}$}}

\def \HI {H{\sc \,i}}

\def\lapp{\ifmmode\stackrel{<}{_{\sim}}\else$\stackrel{<}{_{\sim}}$\fi}
\def\gapp{\ifmmode\stackrel{>}{_{\sim}}\else$\stackrel{>}{_{\sim}}$\fi}

\begin{document}
\def\teff{$T\rm_{eff }$}
\def\kms{$\mathrm {km s}^{-1}$}

\title{
Where is the Cold Neutral Gas in the Hosts of High Redshift AGN?
}

   \subtitle{}

\author{
S. \,J. \,Curran\inst{1},
M.  \,T. \,Whiting\inst{2}
\and J. \,K. \,Webb\inst{1}
}

  \offprints{S. Curran}

\institute{
School of Physics, University of New South Wales, Sydney NSW 2052, Australia
\and
CSIRO Australia Telescope National Facility, PO Box 76, Epping NSW 1710, Australia
\email{sjc@phys.unsw.edu.au}
}

\authorrunning{Curran, Whiting \& Webb}

\titlerunning{Cold Neutral Gas in High Redshift AGN}

\abstract{ Previous surveys for \HI\ 21-cm absorption in $z > 0.1$
  radio galaxies and quasars yield a $\approx40$\% detection rate,
  which is attributed to unified schemes of active galactic nuclei
  (AGN). In this paradigm absorption is only witnessed in (close to)
  type-2 objects, where the central obscuration is viewed (nearly)
  edge-on and thus absorbs the rest frame 1420 MHz emission along our
  sight-line. However, we find this mix of detections and
  non-detections to only apply at low redshift ($z < 1$): From a
  sensitive survey of eight $z\gapp3$ radio sources we find no 21-cm
  absorption, indicating a low abundance of cold neutral gas in (the
  sight-lines searched in) these objects. Analysing the spectral
  energy distributions of these sources, we find that our high
  redshift selection introduces a bias where our sample consists
  exclusively of quasars with ultra-violet luminosities in excess of
  $L_{\rm UV}\sim10^{23}$ W Hz$^{-1}$. This may suggest that we have
  selected a class of particularly UV bright type-1 objects. Whatever
  the cause, it must also be invoked to explain the non-detections in
  an equal number of $z < 0.7$ sources, where we find, for the first
  time, the same exclusive non-detections at $L_{\rm UV}\gapp10^{23}$
  W Hz$^{-1}$. These objects also turn out to be quasars and, 
  from these
  exclusive high UV luminosity--21-cm non-detections, 
  it is
  apparent that orientation effects alone cannot account for the mix
  of 21-cm detections and non-detections at any redshift.

 }
\maketitle{}

\section{Introduction}

Observations of the redshifted 21-cm transition of neutral hydrogen
(\HI) provide a powerful probe of the nature and contents of the early
Universe. For example, this line can be used to:
\begin{enumerate}
\item Measure the baryonic content of the young Universe, when,
  before its consumption by star formation, neutral gas outweighed the
  stars.

  \item Probe the evolution of large-scale structure and galaxy
    formation. At high redshifts, interactions occur more frequently
    and \HI\ observations provide an indispensable means to studying
    the dynamics of galactic mergence and accretion.

    \item Provide a lower limit the time of the epoch of reionisation,
      when neutral hydrogen collapsed forming the first galaxies
      and igniting the first stars.

      \item Measure any changes in the values of the fundamental
        constants of nature: Comparing the redshifted frequencies of
        the 21-cm transition with those of metal-ion and molecular
        lines against laboratory values, can in principle yield
        measures of various combinations of fundamental constants at
        large look-back times (see \citealt{cdk04}).
\end{enumerate}
However, redshifted \HI\ 21-cm absorption systems are currently very
rare with only 67 known at $z\gapp0.1$ (see table 1 of
\citealt{cww+08}). We have therefore embarked upon a large survey to
search for this transition in both systems intervening the
lines-of-sight to background radio sources \citep{cmp+03,ctp+07} and
within the sources themselves (i.e.  for absorption ``associated''
with the host, \citealt{cwm+06,cww+08}). From a recent survey for
associated absorption in $z = 2.9 - 3.8$ radio sources with the
Giant Metrewave Radio Telescope (GMRT), 21-cm was not detected in
any of the eight targets for which good data were obtained. We
discuss the reasons for this here.

\section{Results and Discussion}

\begin{figure*}[t!]
\resizebox{\hsize}{!}{\includegraphics[angle=-90,clip=true]{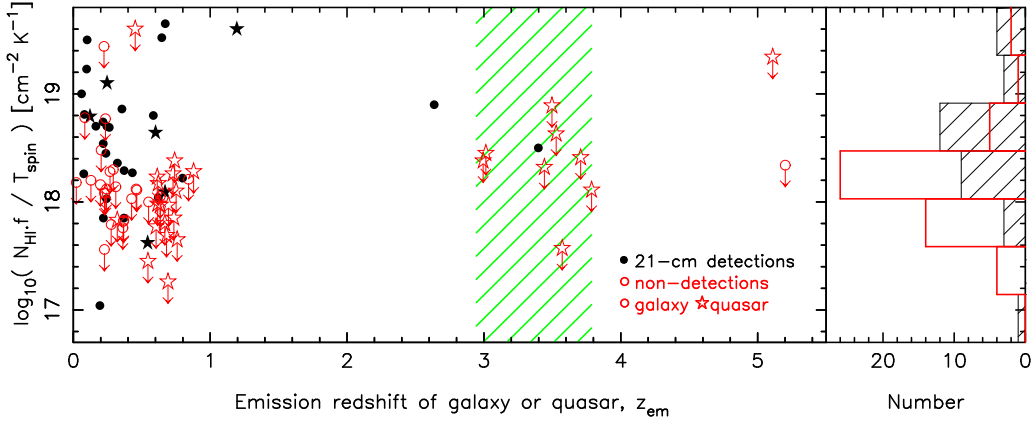}}
\caption{\footnotesize The scaled velocity integrated optical depth of
  the \HI\ line ($1.823\times10^{18}.\int \tau dv$) versus the host
  redshift for the published $z\gapp0.1$ searches for associated 21-cm
  absorption. The filled symbols represent the 21-cm detections and
  the unfilled symbols the non-detections, with stars designating
  quasars and circles galaxies. The hatched region shows the range of our recent
  survey. The detections in this range are from \cite{ubc91,mcm98} and
  the two $z_{\rm em}>5$ non-detections are from \cite{cwh+07}.}
\label{N-z}
\end{figure*}

\begin{figure*}[t!]
\begin{center} 
\includegraphics[scale=0.32]{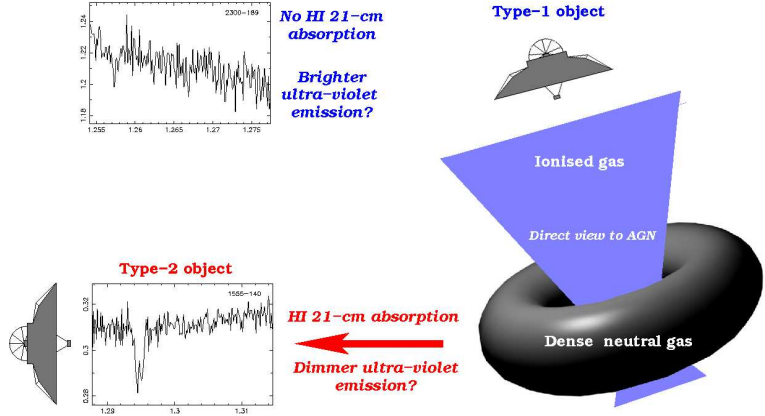}
\end{center} 
\caption{\footnotesize Schematic showing the lines-of-sight to an AGN,
  where according to unified schemes the object type depends upon the
  orientation at which the nucleus is observed: In type-1 objects we
  view the AGN directly and in type-2s this is obscured by a
  circumnuclear torus of dense neutral gas. The 21-cm absorption
  detection shown is in PKS 1555--140, a known type-2 object, and the
  non-detection is in PKS 2300--189, a known type-1 object (the
  spectra are taken from \citealt{cwm+06}, where the ordinate is the
  flux density in Jy and the abscissa the observed frequency in GHz).
}
\label{agn}
\end{figure*}

Our observations are discussed in detail in \citet{cwm+06,cww+08} and
in Fig. \ref{N-z} we show the derived 21-cm line strengths, indicating
the upper limits of our and the previous surveys.
As seen from the figure, our limits are comparable to the vast
majority of detections (which are primarily at low
redshift)\footnote{Note that most of the low redshift non-detections
  have been searched sufficiently deeply to detect 21-cm in most of
  the known absorbers.}. Currently, the known mix of detections and
non-detections (until our survey, mostly at $z\lapp1$) are attributed
to unified schemes of active galactic nuclei (AGN), where the observed
properties of the active galaxy or quasar are due to the orientation
at which the active nucleus is viewed (see \citealt{ant93,up95}): In
type-1 objects, the AGN is viewed directly and in type-2 objects
through a large column of dense obscuring gas, thus giving rise to the
21-cm absorption by the cold neutral gas located along our
line-of-sight (e.g. \citealt{jm94,cb95}, Fig. \ref{agn}).

\begin{figure*}[t!]
\resizebox{\hsize}{!}{\includegraphics[angle=-90,clip=true]{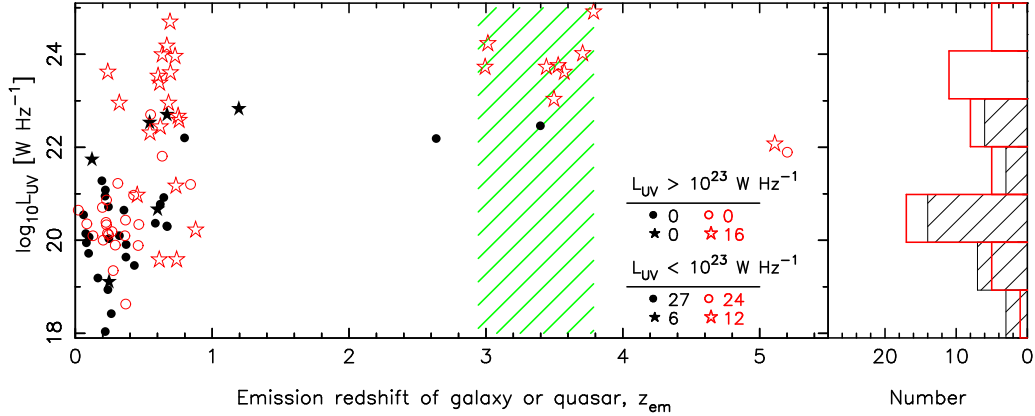}}
\caption{\footnotesize As per Fig. \ref{N-z}, but with the calculated ultra-violet luminosities
on the ordinate. }
\label{lum-z}
\end{figure*}

However, for our sample we obtain exclusive non-detections, where,
according to unified schemes, we may expect the $\approx40$\% (31 out
of 73) mix seen at $z\lapp1$\footnote{Perhaps more, since the density
  of \HI\ at $z \sim 3$ is expected to be higher than it is presently
  (e.g. \citealt{psm+01}).}. With two detections already obtained at
$z = 2.64$ and $3.40$ (Fig. \ref{N-z}), it is clear that our high
redhsift selection (alone, at least) cannot be responsible for the lack
of 21-cm absorption in our targets. If we consider the ultra-violet
($\lambda\sim1000$ \AA) luminosities, however (which we have estimated
from the available optical and near-infrared photometry,
\citealt{cww+08}), we see that all our targets have luminosities of
$L_{\rm UV}\gapp10^{23}$ W Hz$^{-1}$ (Fig. \ref{lum-z}).
Furthermore, although there is a roughly equal mix of detections (33
objects) and non-detections (36 objects) at $L_{\rm UV}\lapp10^{23}$ W
Hz$^{-1}$, the exclusive non-detections at high UV luminosities
relation is also seen for the low redshift sources.

We therefore suspect that the high UV fluxes are ionising the neutral
gas, rendering this undetectable through the 21-cm transition of
hydrogen. This is the first time such a correlation has been noted and,
if indeed the case, perhaps exclusive 21-cm non-detections should have
been expected for our sample, which at luminosity distances of 26 to
34 Gpc are severely flux limited. That is, at such large distances
only the most UV luminous sources are known\footnote{The quasar frame
  UV flux being redshifted into the optical band in these cases.}. It
should be noted, however, that our main selection criterion was
choosing radio-loud sources with $B\gapp19$. This introduces a
luminosity ``ceiling'' of $L_{\rm UV}\lapp3\times10^{24}$ W Hz$^{-1}$,
thus causing us to select the {\it dimmer} ultra-violet sources known
at these redshifts. 

With hindsight, a high redshift selection yielding high luminosity
objects undergoing a large degree of ionisation may be expected,
although, what is surprising is the discovery of the eight 21-cm
non-detections at low redshift (all at $z\leq0.7$, Fig. \ref{lum-z})
for which the UV luminosity also exceeds $L_{\rm UV}\sim10^{23}$ W
Hz$^{-1}$. These objects could be understood in terms of unified
schemes (Fig. \ref{agn}), where we may expect high UV luminosities to
go hand-in-hand with 21-cm non-detections (i.e. all are type-1
objects), if it were not for the non-detections at $L_{\rm
  UV}\lapp10^{23}$ W Hz$^{-1}$ (in fact down to $L_{\rm
  UV}\approx4\times10^{18}$ W Hz$^{-1}$). Again, this is the first
time that such a segregation has been noted and suggests that the
$L_{\rm UV}\gapp10^{23}$ W Hz$^{-1}$ targets (which are all flagged as
quasars) are different from their low luminosity counterparts in which
21-cm also remains undetected (which comprise of a mix of galaxies and quasars).


\section{Summary}

We have undertaken a survey for \HI\ 21-cm absorption in the hosts of
$z = 2.9 - 3.8$ radio sources. Although this has previously been
detected at $z = 2.64$ and $3.40$, we detect no absorption in any of
the eight sources searched. Upon examining the ultra-violet
luminosities of our targets, we find that all have $L_{\rm
  UV}\gapp10^{23}$ W Hz$^{-1}$ (which the two high redshift detections
do not), a relation we find also to apply to the low redshift searches
in the literature. This suggests that the high ultra-violet fluxes are
ionising the hydrogen to neutral column densities below our detection
threshold. If this is only occuring along our line-of-sight to the
active nucleus, our findings are consistent with these being
exclusively type-1 objects (all are flagged as quasars according to
their morphologies, although there are quasars detected in
21-cm). However, the fact remains that there exists a roughly equal
mix of 21-cm detections and non-detections below $L_{\rm
  UV}\sim10^{23}$ W Hz$^{-1}$ (which are a mix of quasars and galaxies).
The orientation effects invoked by unified schemes of active galactic
nuclei are generally believed to be responsible for this mix, although
our result would suggest that there are both low and high UV-luminous
type-1 objects, while type-2 objects exclusively have $L_{\rm
  UV}\lapp10^{23}$ W Hz$^{-1}$.



\end{document}